\documentclass[12 pt]{article}
\usepackage{epsfig}
\usepackage{graphicx}
\title{Coupled anharmonic
oscillators: the Rayleigh-Ritz approach versus the collocation approach}
\author{Arkadiusz Kuro\'s and Anna Okopi\'nska\\
Institute of Physics, Jan Kochanowski University,\\
\'Swi\c{e}tokrzyska 15, 25-406 Kielce, Poland}
\date{}
\begin{document}
\maketitle
\begin{abstract}
\noindent For a system of coupled anharmonic oscillators we compare the convergence rate of the variational collocation approach presented recently by Amore and Fern\'andez (2010 Phys.Scr.81 045011) with the one obtained using the optimized Rayleigh-Ritz (RR) method. The monotonic convergence of the RR method allows us to obtain more accurate results at a lower computational cost.
\end{abstract}

\section{Introduction}
Since the 1930 s, the linear Rayleigh-Ritz (RR) variational method~\cite{RS} has been the primary tool for determining the low energy part of the spectrum of quantum systems. In this approach, the trial function is expanded into a finite set of basis functions, and the coefficients of the expansion are determined by the variational principle. The approximations provided by the eigenvalues of the RR matrix are upper bounds that approach the exact energies monotonically from above on increasing the number of included functions of the basis~\cite{RS}. However, the matrix elements are given by integrals  and the number of the basis functions required to obtain accurate results grows exponentially with the number of degrees of freedom, which renders the method computationally prohibitive for complex systems. There are basically two strategies to reduce the computational cost of the method: choosing the basis functions in a sophisticated way or using approximations in calculating the Hamiltonian matrix elements. Among various implementations of the first strategy 
the introduction of nonlinear parameters into the functions of the basis is one of the most effective~\cite{Hyl}. The adjustable parameters may be fixed so as to improve the convergence for a particular state~\cite{rych} or a group of states~\cite{ao2}. An example of the second strategy is the collocation method where the matrix elements are approximated by $N-$point quadratures. The recently published paper on "Variational collocation for systems of coupled anharmonic oscillators"~\cite{coll} combines both strategies mentioned above, demonstrating that by introducing adjustable parameters the convergence of the collocation method may be improved. However, it should be remembered that in contrast to the RR method, the collocation method does not have a variational character. Discretization in the collocation approach introduces approximation in the Hamiltonian matrix, as a consequence of which the results do not represent upper bounds to the exact bound-state energies. Introducing the adjustable parameters does not change this fact; therefore the method proposed by Amore and Fernandez should be termed optimized rather than variational. 
\section{The model}
To illustrate the similarities and differences between optimization in the RR and collocation methods, we consider the example of the Pullen-Edmonds Hamiltonian
\begin{equation}
    \hat{H}=-\frac{\partial^2}{\partial x^2}-\frac{\partial^2}{\partial y^2}+ x^2 +y^2+\lambda x^2  y^2
    \label{eq:hdim2}
\end{equation}
that has been studied in Ref.~\cite{coll}. The parameter $\lambda$ represents the coupling between the two degrees of freedom. Amore and Fernandez observed that introducing a rotation parameter, the convergence of the collocation method may be improved further. The optimal rotation angle has been found to be close to $\pi/4$, which corresponds to the Hamiltonian
\begin{equation}
    \hat{H}=-\frac{\partial^2}{\partial x^2}-\frac{\partial^2}{\partial y^2}+ x^2 +y^2+\frac{\lambda}{4} (x^2 - y^2)^2.
    \label{eq:hdim2ROT}
\end{equation} Hamiltonians (\ref{eq:hdim2}) and (\ref{eq:hdim2ROT}) have the same spectrum but the speed of convergence of the collocation method turns to be considerably higher in case of the application to the rotated one (\ref{eq:hdim2ROT}).

Below we consider the application of the RR method to both Hamiltonians using basis functions $\phi_{nm}^{\alpha}(x,y)=\phi_{n}^{\alpha}(x)\phi_{m}^{\alpha}(y)$ constructed as tensor products of the functions $\phi_{n}^{\alpha}(x)$ from a basis in one-dimensional (1D) space that depends on an arbitrary parameter $\alpha$. The $M$th order approximations are obtained with the use of $M$ even-parity 1D functions, which corresponds to diagonalization of the RR matrix of dimension $M^2$. Optimization is performed by choosing the value of $\alpha$ so as to make the trace of that matrix stationary at each order of approximation~\cite{ao2}. In our procedure, we take the advantage of symbolic calculation in the Mathematica package by performing the numerical algebra exactly in terms of the ratio of integers, which allows determination of the eigenvalues to arbitrary accuracy.
\section{Trigonometric basis}
First we calculate the accurate values of ground-state energy of the coupled oscillators through the optimized RR method, using the basis of trigonometric functions
\begin{equation}
\phi_{n}^{L}(x)=\frac{1}{\sqrt L}Cos\left[\left(n+\frac{1}{2}\right)\frac{\pi x}{L} \right],
\label{trig}
\end{equation}
which are eigenfunctions of the infinite potential well of a half-width $L$ that is taken as a nonlinear parameter. The same basis has been used by Amore and Fernandez in their collocation approach~\cite{coll}. In table~$\ref{eq:tabL}$, we compare the approximations to the ground-state energy $\varepsilon_0(1)$ and $\varepsilon_0(2)$ for the Hamiltonian (\ref{eq:hdim2}) and (\ref{eq:hdim2ROT}), respectively, for different values of $M$. The optimal value of the nonlinear parameter $L_{opt}$ that renders the trace of the given order RR matrix stationary is also shown. The underlined digits are those that coincide with the exact result. We observe that both the series of approximations converge monotonically from above to the exact value with increasing $M$, but the convergence for the Hamiltonian (\ref{eq:hdim2ROT}) is much quicker. This is similarly true at other couplings as can be seen in table \ref{eq:tabLcomp}, where the results obtained with $M=35$ are compared for different values of $\lambda$. We conclude that the same rotation that speeds up the convergence of the collocation method improves also the convergence of the RR method.
\begin{table} [!ht]
\begin{center}
\small
\begin{tabular}{|c||c|c||c|c|}
\hline
$M$ &$L_{opt}(1)$& $\varepsilon_0(1)$ &$L_{opt}(2)$& $\varepsilon_0(2)$ \\
\hline
10&2.627&\underline{3.019}70464&3.015&\underline{3.0191777}2606335945122\\
15&2.988&\underline{3.019}21402&3.443&\underline{3.0191777147}8547514131\\
20&3.278&\underline{3.0191}8109&3.787&\underline{3.01917771477}200287135\\
25&3.525&\underline{3.01917}810&4.077&\underline{3.019177714771967}54393\\
30&3.741&\underline{3.0191777}7&4.332&\underline{3.01917771477196738}793\\
35&3.935&\underline{3.0191777}2&4.560&\underline{3.0191777147719673869}2\\
\hline
\end{tabular}
\caption{RR approximations to the ground-state energy of Hamiltonians (\ref{eq:hdim2}) and (\ref{eq:hdim2ROT}) with $\lambda=10$ obtained using
$M$ functions from the trigonometric basis with optimized quarter period $L_{opt}$. The underlined digits are those that coincide with the exact result.}
\label{eq:tabL}
\end{center}
\end{table}

\begin{table} [!ht]
\begin{center}
\small
\begin{tabular}{|r||c|c||c|c|}
\hline
$\lambda$ &$L_{opt}(1)$ & $\varepsilon_0(1)$ & $L_{opt}(2)$ &  $\varepsilon_0(2)$ \\
\hline
5   & 4.41 & \underline{2.65390977}8 & 5.10&  \underline{2.6539097779532153534905}8\\
10  & 3.93 & \underline{3.0191777}23 & 4.56 &\underline{3.0191777147719673869}2058\\
100 & 2.69 & \underline{5.46097}1628 & 3.12 & \underline{5.46097039792335}813076119\\
1000& 3.66 & \underline{11.2324}4729 & 2.13 & \underline{11.2324392672}100309737449\\
10000& 1.83& \underline{23.94}601185 & 1.47 & \underline{23.94598962781}97863410786\\
\hline
\end{tabular}
\caption{RR approximations to the ground state-energy of Hamiltonians (\ref{eq:hdim2}) and (\ref{eq:hdim2ROT}) for several values of $\lambda$ obtained using
$M=35$ functions from the trigonometric basis with optimized quarter period $L_{opt}$.}
\label{eq:tabLcomp}
\end{center}
\end{table}
\section{Harmonic oscillator basis}
For comparison, we applied the optimized RR method using a different basis, namely the set of eigenfunctions of the harmonic oscillator with an arbitrary frequency $\Omega$ that are given by
\begin{equation}
\phi_{n}^{\Omega}(x)= \left(\frac{\Omega}{\pi}\right)^{1/4} \frac{1}{\sqrt{ 2^{n}n!}} H_n \left(\sqrt{\Omega} x\right) e^{-\frac{\Omega x^2}{2}} .
\label{ho}
\end{equation}
The approximations to ground-state energy obtained with the harmonic oscillator basis are presented in tables $\ref{eq:tabO}$ and $\ref{eq:tabOcomp}$. Comparison with the results obtained with the trigonometric basis in tables $\ref{eq:tabL}$ and $\ref{eq:tabLcomp}$ shows that the convergence is much faster for the harmonic oscillator eigenfunctions, which may be attributed to the fact that they fulfill the proper boundary conditions at infinity. The convergence for the rotated Hamiltonian (\ref{eq:hdim2ROT}) is much better than for the original one (\ref{eq:hdim2}) and the obtained results agree with the best available calculations of the perturbative-variational method performed to an accuracy of 15 digits~\cite{Saav}. For Hamiltonian (\ref{eq:hdim2ROT}) with $\lambda=5$ the precision of our calculations with $M=35$ harmonic oscillator functions reaches 30 digits. For larger values of $\lambda$, a similarly high precision can be easily obtained with reasonably increased $M$, which would be not an easy task for the original Hamiltonian (\ref{eq:hdim2}).
\begin{table} [!ht]
\begin{center}
\small
\begin{tabular}{|c||c|c||c|c|}
\hline
$M$ &$\Omega_{opt}(1)$& $\varepsilon_0(1)$ &$\Omega_{opt}(2)$& $\varepsilon_0(2)$ \\
\hline
5  &3.65& \underline{3.0}2083109727&3.04& \underline{3.019}2115512360457661267360\\
10 &4.64& \underline{3.019}20677896&3.77& \underline{3.01917771}51455247319696151\\
15 &5.32& \underline{3.01917}868488&4.30& \underline{3.0191777147719}932650455193\\
20 &5.86& \underline{3.0191777}6226&4.71& \underline{3.0191777147719673}955033821\\
25 &6.31& \underline{3.01917771}779&5.07& \underline{3.01917771477196738691}95328\\
30 &6.71& \underline{3.01917771}501&5.38& \underline{3.0191777147719673869116}933\\
35 &7.06& \underline{3.0191777147}9&5.65& \underline{3.019177714771967386911678}9\\
\hline
\end{tabular}
\caption{RR approximations to the ground-state energy of Hamiltonians (\ref{eq:hdim2}) and (\ref{eq:hdim2ROT}) with $\lambda=10$ obtained using
$M$ functions from the harmonic oscillator basis with optimized frequency $\Omega_{opt}$.}
\label{eq:tabO}
\end{center}
\end{table}

\begin{table} [!ht]
\begin{center}
\small
\begin{tabular}{|r||c|c||c|c|}
\hline
$\lambda$ &$\Omega_{opt}(1)$ & $\varepsilon_0(1)$ & $\Omega_{opt}(2)$ &  $\varepsilon_0(2)$ \\
\hline
5    & 5.63&\underline{2.653909777953}6& 4.51&\underline{2.6539097779532153534905698061}7\\
10   & 7.06&\underline{3.0191777147}930& 5.65&\underline{3.019177714771967386911678}93635\\
100  &15.13&\underline{5.460970}4165739&12.08&\underline{5.46097039792335524}182772613188\\
1000 &32.56&\underline{11.232439}458359&25.98&\underline{11.232439267209851}6856244029369\\
10000&70.14&\underline{23.9459}90215098&55.94&\underline{23.94598962781893}96640103254023\\
\hline
\end{tabular}
\caption{RR approximations to the ground state-energy of Hamiltonians (\ref{eq:hdim2}) and (\ref{eq:hdim2ROT}) for several values of $\lambda$ obtained using
$M=35$ functions from the harmonic oscillator basis with optimized frequency $\Omega_{opt}$.}
\label{eq:tabOcomp}
\end{center}
\end{table}

\section{Comparison of the results and conclusion}
\begin{figure} [h!]
\begin{center}
\includegraphics{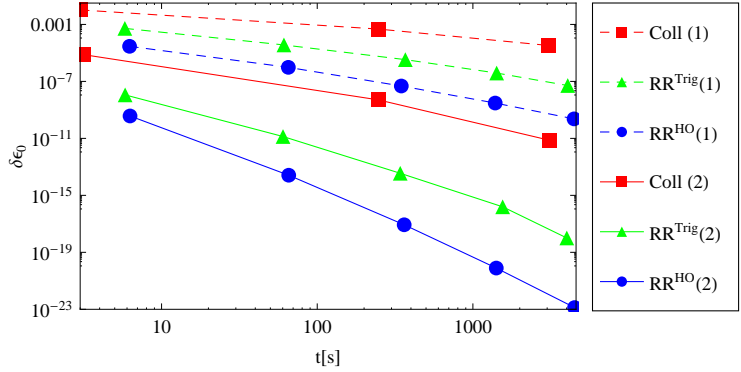}
 \caption{Precision achieved by the optimized collocation (COLL) and RR method with trigonometric (TRIG) and harmonic oscillator (HO) eigenfunctions for the Hamiltonians (\ref{eq:hdim2}) and (\ref{eq:hdim2ROT}) with $\lambda=10$ as functions of CPU time.}
 \label{eq:p_nr1}
 \end{center}
\end{figure}

In this paper, we studied the convergence of the optimized RR method for the ground-state energy of the coupled anharmonic oscillators using two basis sets: trigonometric functions and harmonic oscillator eigenfunctions. The comparison of the effectiveness of the methods with that of the optimized collocation method~\cite{coll} is summarized in figure \ref{eq:p_nr1} by plotting the precision achieved in the computation performed with Mathematica ver. 7.0. The precision calculated as a difference between the approximate and exact result $\delta \epsilon_{0}$ is shown in a log-log plot as a function of CPU time. An important observation is that the rotation that has been shown to improve the convergence of the collocation method~\cite{coll} accelerates also the convergence of both variants of the RR method. It is seen that far greater accuracy in the ground-state energy is obtained by using the rotated Hamiltonian (\ref{eq:hdim2ROT}) than by increasing the size of the basis. The optimized RR method is more effective with the harmonic oscillator basis than with the trigonometric one, but in both cases the convergence is quicker than that of the optimized collocation approach. This is opposed to the common understanding that the computational cost of the collocation method is less then that of the RR method~\cite{comp}. Our results show that the RR method with an appropriately optimized basis offers a similar accuracy to the optimized collocation method with much lower computational cost.

Another important advantage of the RR method is its variational character, which allows us to state that the lower the achieved result, the closer it is to the exact one. Discretization in the collocation approach facilitates calculation of the Hamiltonian matrix but destroys the variational behavior. Normally the monotonic convergence from above is recovered when the number of grid point is sufficiently large but this is not automatically granted. For instance we observe that the approximations obtained by the collocation method from the rotated Hamiltonian (\ref{eq:hdim2ROT}) in table 4 of ~\cite{coll} are below the accurate ground-state energies (~\cite{Saav} and our table \ref{eq:tabOcomp}) and they decrease with increasing $M$, which suggests that they do not converge to the proper result. This is a warning that when using the collocation method one has to be very cautious about deriving accurate results.

We believe that our conclusions are relevant not only for the basis sets considered in this work but also for other types of basis that are used in higher-dimensional problems.


\begin{thebibliography}{9}
\bibitem{RS} Reed M and Simmons B 1978\emph{Methods of Modern
Mathematical Physics, Vol.4} (Academic, New York) p.82
\bibitem{Hyl} Hylleraas E A, Zeit.Phys.\emph{54} 347 (1929)
\bibitem{rych} Cencek W, Komasa J, and Rychlewski J, in
\emph{Handbook on Parallel
and Distributed Processing} (Springer, Berlin, 2000), p. 505.
\bibitem{ao2} Okopi\'nska A 1987 \emph{Phys.Rev.D} {\bf 36} 1273
\bibitem{coll} Amore P and Fern\'andez F M 2010 \emph{Phys.Scr.} {\bf 81} 045011
\bibitem{Saav} Arias de Saavendra  F and Buendia E 1991 \emph{J.Phys.A:Math.Gen.} {\bf 24} L1209
\bibitem{comp} Yang W and Peet A C, 1988 \emph{Chem. Phys. Lett.} {\bf 153} 98.
\end{thebibliography}
\end{document}